\documentclass[conference]{IEEEtran}
\IEEEoverridecommandlockouts

\usepackage{amsfonts,
			amsmath,
			booktabs,
			graphicx,
			subfigure,
			url,
			algorithm,
			algpseudocode,
            ulem,
			pgfplots,
			xcolor}

\pgfplotsset{compat=1.18}

\newif\ifanonymous
\anonymousfalse  

\begin{document}

\title{Numerical Kernels on a Spatial Accelerator: A Study of Tenstorrent Wormhole

	\thanks{
		This material is based upon work supported by the U.S. Department of Energy, Office of Science, Office of Advanced Scientific Computing Research, Department of Energy Computational Science Graduate Fellowship under Award Number DE-SC0025528.


	}
}


\makeatletter
\newcommand{\linebreakand}{%
	\end{@IEEEauthorhalign}
	\hfill\mbox{}\par
	\mbox{}\hfill\begin{@IEEEauthorhalign}
}
\makeatother

\author{\IEEEauthorblockN{Maya Taylor}
	\IEEEauthorblockA{\textit{Siebel School of Computing and Data Science} \\
		\textit{University of Illinois Urbana-Champaign}\\
		Urbana, IL, USA \\
		mayat4@illinois.edu}
	\and
	\IEEEauthorblockN{Carl Pearson, Luc Berger-Vergiat}
	\IEEEauthorblockA{\textit{Center for Computing Research} \\
		\textit{Sandia National Laboratories}\\
		Albuquerque, NM, USA \\
		cwpears@sandia.gov, lberge@sandia.gov}
	\linebreakand 
	\IEEEauthorblockN{Giovanni Long}
	\IEEEauthorblockA{\textit{Computer Science Department} \\
		\textit{University of California Santa Barbara}\\
		Santa Barbara, CA, USA \\
		giovanni\_long@ucsb.edu}
	\and
	\IEEEauthorblockN{Jan Ciesko}
	\IEEEauthorblockA{\textit{Center for Computing Research} \\
		\textit{Sandia National Laboratories}\\
		Albuquerque, NM, USA \\
		jciesko@sandia.gov}
}
\maketitle

\begin{abstract}
	As AI accelerators gain prominence, their potential for traditional scientific computing workloads remains unclear.
	This paper explores Tenstorrent's Wormhole architecture, a spatial computing platform designed for neural network acceleration, by implementing three numerical kernels and composing them into a conjugate gradient solver.
	We present architecture-specific optimizations for sparse numerical algorithms, evaluate their performance against Nvidia GPUs, and expose both challenges and opportunities in porting numerical methods to spatial architectures.
	Most notably, a single Wormhole die achieves preconditioned conjugate gradient performance within $8.8\times$ of an Nvidia H100 despite older 12nm process technology and substantially lower memory bandwidth. These results motivate further study of AI accelerators for workloads traditionally dominated by CPUs and GPUs.
\end{abstract}


\maketitle

\section{Introduction}

The advent of compute-intensive neural networks has driven proliferation of novel computer architectures designed to accelerate training and inference, including products developed by Cerebras, Graph Core, Groq, SambaNova, and Tenstorrent, among others.
These architectures broadly feature small, simple processing elements (PEs) with single-instruction multiple-data (SIMD) parallelism for high-throughput math,
small scratchpad memories, and a user-controlled network-on-chip (NoC).
Often called spatial or dataflow architectures, they emphasize the increased importance of explicit data placement and movement. They are particularly efficient for dense tensor work and have also shown promise for traditional modeling and simulation (modsim) workloads~\cite{rocki2020faststencilcodecomputationwaferscale, santos2024breaking}.

Despite these successes, the suitability of spatial architectures for traditional numerical methods remains unclear. We investigate this question through a case study on Tenstorrent’s Wormhole, implementing three numerical kernels and a full conjugate gradient solver.

\begin{enumerate}
	\item Implementing and evaluating three key numerical kernels on Wormhole: basic vector arithmetic, global reduction, and stencil computations.
	\item Demonstrating a modsim-relevant numerical algorithm (preconditioned conjugate gradient) at 32-bit precision on Wormhole for the first time.
	\item Most notably, we show that our solver runs within 4.3× (16-bit) and 8.8× (32-bit) of an Nvidia H100 despite Wormhole being built in an older 12nm process and offering much lower memory bandwidth.
\end{enumerate}


\section{Background and Related Work}
\label{sec:background}

The work presented in this paper builds primarily on Brown and Barton~\cite{brown2024accelerating}, which explores DDR access patterns and energy efficiency for a Jacobi iterative method on Tenstorrent's earlier ``Grayskull'' architecture.
Grayskull and Wormhole share many architectural features, but Wormhole adds support for additional datatypes (including FP32), higher memory capacity, and optimizations for multi-device scaling~\cite{doerner2024analysis}.

Our work advances this earlier implementation with a more demanding iterative method, a higher-dimensional stencil, and on-chip SRAM staging.
Beyond Tenstorrent, most prior work on modsim algorithms for spatial dataflow architectures targets Cerebras' Wafer-Scale Engines (WSE) ~\cite{santos2024breaking, sai2024matrix, tramm2024efficient}.
Other work explores dataflow architectures for stencils and basic collective primatives
~\cite{de2021stencilflow, schnyder20242d, sai2024automated, luczynski_2024}, specialized sparse solver hardware ~\cite{feldman2024azul}, and relevant programming abstractions and compilation techniques ~\cite{woo2022disruptive,van2025system}.

\section{Tenstorrent Wormhole}
\label{sec:wormhole}

The Tenstorrent n300d is a PCIe card featuring two ``Tensix'' Wormhole accelerator dies.
The PCIe card is equipped with 24GB of GDDR6 DRAM accessible by both Tensix dies.
Each Tensix die is composed of a 2D grid
of 10x12 compute elements, of which 80 are designated for compute (``Tensix cores''), with the remainder allocated to DRAM, PCIe, ARC, or Ethernet management.
In addition, the chip has a 2D Torus NoC for data movement between cores and DRAM.
The programmer interacts primarily with memory movement cores and Tensix cores, which are the focus of the remainder of this section.

\begin{figure}
	\centering
	\includegraphics[width=.4\textwidth]
	{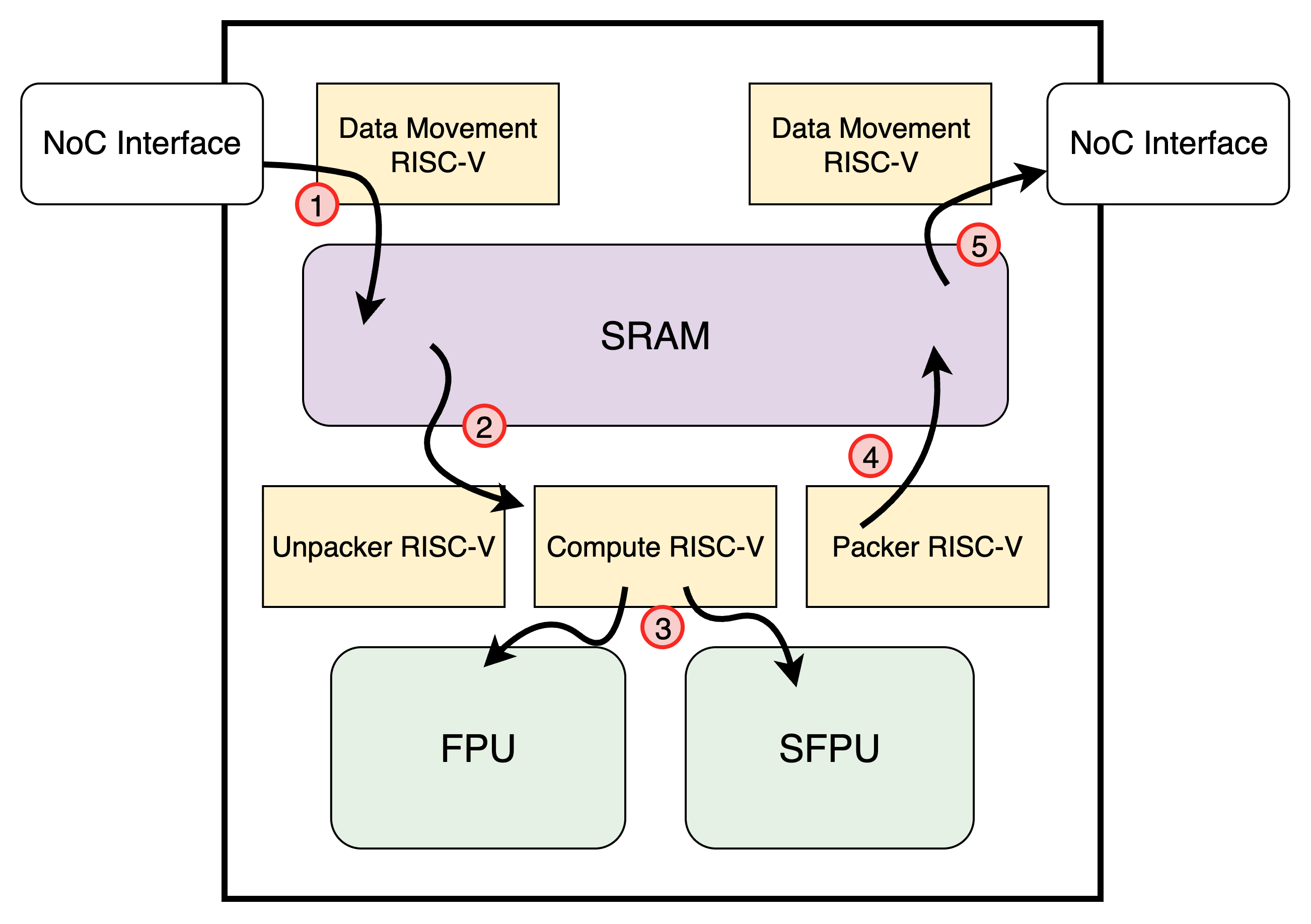}
	\caption{Block diagram of a Tensix core.
		Two of the baby RISC-V cores are connected to the NoC and manage data movement to/from DRAM and other cores. The other three baby RISC-V cores manage data movement
		to/from SRAM and the compute units. Typical dataflow is as follows: first (1) data is moved from the NoC to SRAM by one of the data movement cores. Then, (2) it is moved to registers by the unpacker where (3) the compute core prompts computation on one of the compute units. (4) The result is moved from registers to SRAM by the packer where (5) data can be moved from SRAM to the NoC by the second data movement core.
	}
	\label{fig:tensix-core}
\end{figure}

Figure \ref{fig:tensix-core} illustrates the architecture at a high level, with an outline of a typical data flow through a Tensix core.
To program the Wormhole device, we use the \emph{tt-metal} API, which provides low-level access to the Tensix cores and DRAM.
The programmer manages data movement to/from DRAM, between cores, and to/from the compute units.
Similarly to existing accelerator models, tt-metal programs require a C++ host program
which runs on the CPU and coordinates memory movement to and from the device, as well as kernel launches.


\subsection{Tiles and Circular Buffers}
The most central component of the tt-metal API is the \textit{tile abstraction}, which dictates data movement and computation patterns. Tiles are 2D arrays of 1024 elements, and all communication and computation is performed at the tile granularity. For dense matrix operations, this abstraction is often very practical, but for more general numerical workloads it poses challenges, some of which are discussed below.





Data movement of tiles is regulated through the use of ``circular buffers'', which are FIFO queues statically allocated in SRAM and used to stage data for movement between cores and registers.
These circular buffers facilitate pipelining of all communication and act as a synchronization mechanisms between the 5 baby RISC-V cores, encouraging an asynchronous, dataflow programming style within a Tensix core.
In the most common use case, one circular buffer is reserved for movement of tiles from SRAM to the FPU (matrix unit) or SFPU (vector unit) source registers, and a second is reserved for movement from the destination registers back to SRAM.

Each Tensix core has two source registers (SrcA and SrcB) and one destination register set (Dst) which are used to move data to and from the FPU or SFPU, as described below. The registers enforce the tile abstraction in hardware as each register is composed of 64 rows of 16 elements.
Data movement is also restricted to tile and sub-tile granularity: reads from DRAM must be 32B aligned, and L1 access and writes to DRAM must be 16B aligned.

\subsection{Compute Units}


\textbf{The FPU} is designed to operate in a SPMD fashion only on sub-tiles of 8x16 16-bit elements.
Table \ref{tab:tile-properties} summarizes the single-cycle capabilities of the FPU, demonstrating its utility for matrix multiplication in particular. While the FPU is designed for such matrix operations, it also supports element-wise arithmetic and reductions on tiles, but it is restricted to a maximum of 19-bit data formats (for our purposes, we consider only BF16 precision).

\begin{table}[h]
	\centering
	\begin{tabular}{|c|c|}
		\hline
		Operation                & Size                \\
		\hline
		Matrix Multiply          & 8x16 x 16x16 = 8x16 \\
		Reduction                & 16x16               \\
		Element-wise Add/Sub/Mul & 8x16                \\
		\hline
	\end{tabular}
	\caption{Single-cycle capabilities of the Wormhole FPU.}
	\label{tab:tile-properties}
\end{table}

\textbf{The SFPU,} in comparison, is designed for vector operations on tiles, and can perform basic element-wise arithmetic as well as more complicated functions (like exponentiation and square root). It supports both 16-bit and 32-bit formats and can be considered a traditional SIMD unit with 32 lanes of 32 bits per lane, requiring 2 cycles for most element-wise arithmetic operations on 64 16-bit elements, or 32 16-bit elements per cycle.

Both the FPU and SFPU require unpacking from SRAM to the source registers and subsequent packing from the destination registers back to SRAM, but the SFPU additionally requires load-store operations to move data from the registers to the vector lanes and back. For this reason, in addition to the lower compute throughput, the SFPU is substantially more expensive than the FPU.

\section{Element-wise Arithmetic}
\label{sec:arithmetic}
Both the SFPU and FPU can perform basic element-wise arithmetic operations on tiles, including addition, subtraction, and multiplication.
As discussed above, the throughput of these operations differs between the two compute units, with the FPU generally achieving 128 operations
per clock cycle limited to 19-bit data formats, while the SFPU supports both 16-bit and 32-bit formats and achieves 32 and 16 operations per clock cycle respectively.

\begin{figure}[b]
	\centering
	\resizebox{0.35\textwidth}{!}{
		\begin{tikzpicture}
    \begin{axis}[
            xlabel={arithmetic intensity (FLOPs/Byte)},
            ylabel={performance (FLOPs/cycle)},
            xmode=log,
            ymode=log,
            grid=major,
            xmin=0.01, xmax=10,
            ymin=1, ymax=1000,
        ]

        \addplot[only marks, mark=*, mark size=2pt, blue]
        table[col sep=comma, x=ai, y=perf] {MayaTaylor_Figs/vector_addition/add.csv};
        \addlegendentry{FPU Vector Addition}

        \addplot[only marks, mark=square*, mark size=2pt,red]
        table[col sep=comma, x=ai, y=perf] {MayaTaylor_Figs/vector_addition/sfpu_add.csv};
        \addlegendentry{SFPU Vector Addition}

        \addplot[domain=0.01:2, cyan!70!blue, ultra thick, samples=2] {64*x}
        node[pos=.5, above, sloped] {Pack/Unpack Memory BW}; 

        \addplot[domain=0.01:1, red!80!black, ultra thick, samples=2] {32*x}
        node[pos=.7, below, sloped] {Copy Memory BW }; 

        \addplot[domain=2:100, cyan!70!blue, ultra thick, samples=2] {128}
        node[pos=-1, above] {FPU Peak FLOPs}; 

        \addplot[domain=1:100, red!80!black, ultra thick, samples=2] {32}
        node[pos=-.7, above] {SFPU Peak FLOPs}; 

        \addplot[domain=0.01:1, red!80!black, dotted, thick, samples=2] {32};
        \addplot[domain=.01:2, cyan!70!blue, dotted, thick, samples=2] {128};
        \addplot[domain=2:100, cyan!70!blue, dotted, thick, samples=2] {64*x};
        \addplot[domain=1:100, red!80!black, dotted, thick, samples=2] {32*x};

    \end{axis}
\end{tikzpicture}
	}
	\caption{Roofline model for Wormhole Tensix architecture for 16-bit element-wise addition, with implementation variants for FPU and SFPU. Both data points represent runs with 256 tiles per Tensix core, or 262,144 elements.}
	\label{fig:roofline}
\end{figure}
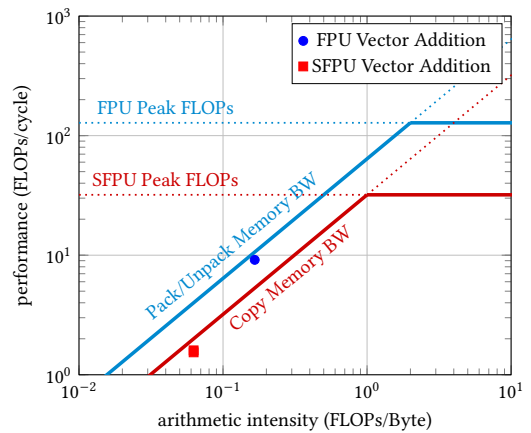

Figure \ref{fig:roofline} presents a roofline model for the Wormhole Tensix architecture specific to 16-bit element-wise addition, where we consider only a single Tensix core and its local SRAM. The memory bandwidth is calculated based on the peak throughput of the packer and unpacker (64B/clk), as they move tiles from SRAM to registers and vice versa.
In this benchmark, data is streamed in from DRAM via the NoC, the core performs vector addition, and the result is written back to DRAM.
For the FPU implementation, the arithmetic intensity (AI) is computed as expected: 1 FLOP per 6 bytes moved. The FPU achieves near-peak performance.

The SFPU implementation is about 6× slower than the FPU for the same operation, due to its additional Dst-register copy (32B/cycle) and load-store overhead to move data to and from the vector lanes, giving an effective AI of 1 FLOP per 16 bytes moved (Figure \ref{fig:roofline}).


Due to the increased cost of SFPU operations, and therefore increased cost of any FP32 arithmetic on Wormhole,
there is a strong incentive to use the FPU and half precision whenever possible. We explore both BF16 and FP32 in this work
but note that the Wormhole is most suitable for BF16 workloads.

\section{Global Reduction}
\label{sec:reduction}

To explore global collective operations on the Wormhole architecture, we consider a dot product. This operation is essential for many numerical algorithms and provides a useful benchmark for evaluating the performance of collective communication on the Wormhole architecture.

The dot product takes two input vectors, distributed equally throughout the Wormhole grid, and involves element-wise multiplication of both input vectors followed by a global sum reduction to produce a single scalar result. Every Tensix core begins with a subset of tiles of both input vectors and computes the element-wise multiplication of tile pairs and then a sum of all local tiles to produce a single tile of output. This single tile can then be reduced to a scalar value via the FPU reduction.



We compare two reduction granularities: reducing each tile to a scalar locally before sending (Method 1) versus sending full tiles and reducing only at the root (Method 2). The two perform similarly, with Method 1 giving a marginal 1.8\% speedup at the largest grid size — unsurprising given the NoC's low latency and the high relative cost of SFPU arithmetic.

Another design decision lies in how to route the global reduction through the Wormhole NoC. We explore two variants of routing patterns. In both, partial sums are sent through the NoC and reduced further at every hop. At every core, only a single partial result is sent onward.
\begin{itemize}
	\item Route 1: send data leftward across all rows of cores and then upward to the top left core. This implementation has the advantage that Tensix cores must handle at most 2 incoming tiles, simplifying the routing logic, but doesn't maximize parallel usage of the NoC.
	\item Route 2: send data to the center of the grid, minimizing the distance traveled. This implementation is more complicated and requires that the center core handle 4 incoming tiles, but achieves better parallel usage of the NoC.
\end{itemize}



\begin{figure}[tb]
	\centering
	\resizebox{0.35\textwidth}{!}{
		\begin{tikzpicture}
  \begin{axis}[
      xlabel={number of cores},
      ylabel={speedup (\%)},
      legend pos=north west,
      grid=major,
      xmin=-2, xmax=52,
      mark size=2pt,
    ]

    \addplot[color=blue, mark=*, solid] table[x expr=(\thisrow{kernel_size}*\thisrow{kernel_size}), y=percent_improvement, col sep=comma] {MayaTaylor_Figs/reduction_pattern/speedup_1.csv};
    \addlegendentry{1 tile}

    \addplot[color=cyan, mark=square*, solid] table[x expr=(\thisrow{kernel_size}*\thisrow{kernel_size}), y=percent_improvement, col sep=comma] {MayaTaylor_Figs/reduction_pattern/speedup_4.csv};
    \addlegendentry{4 tiles}

    \addplot[color=orange, mark=triangle*, solid] table[x expr=(\thisrow{kernel_size}*\thisrow{kernel_size}), y=percent_improvement, col sep=comma] {MayaTaylor_Figs/reduction_pattern/speedup_64.csv};
    \addlegendentry{64 tiles}

    \addplot[color=green!60!black, mark=diamond*, solid] table[x expr=(\thisrow{kernel_size}*\thisrow{kernel_size}), y=percent_improvement, col sep=comma] {MayaTaylor_Figs/reduction_pattern/speedup_128.csv};
    \addlegendentry{128 tiles}

    \addplot[color=black, dashed, line width=1pt, domain=0.5:50] {0};

  \end{axis}
\end{tikzpicture}
	}
	\caption{Analysis of speedup achieved with Route 2 over Route 1. Results show time per iteration averaged over 100 iterations.}
	\label{fig:reduction-patterns-speedup}
\end{figure}
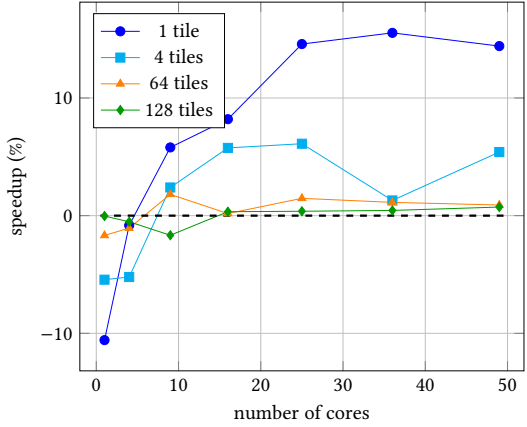

Figure \ref{fig:reduction-patterns-speedup} compares the two routing patterns as the problem size scales using method 2 of partial result granularity (reducing to a scalar only at the end). The results demonstrate that the expected advantage of Route 2 is minimal, especially as the problem size increases and the local compute dominates. Route 2 achieves around a 15\% speedup when using a single tile per core, but at 128 tiles per core the speedup is negligible. To the left of Figure \ref{fig:reduction-patterns-speedup}, the speedup is negative, due to the increased complexity of Route 2 outweighing the benefits. Overall, these results suggest that while Route 2 can provide some performance benefits at small problem sizes, the network is so low latency that simpler routing is sufficient for larger problem sizes.

\section{Stencil Collectives}
\label{sec:stencil}

The third numerical kernel we consider is a 7-point 3D stencil computation. Stencils are widely used in scientific computing applications, particularly for solving differential equations using finite difference methods. They involve updating each point in a grid based on a weighted combination of its neighboring points. The 7-point stencil specifically refers to a 3D grid where each point is updated based on its six immediate neighbors (left, right, up, down, front, back) and itself.
Our contribution in this section focuses on adapting the stencil computation to the Wormhole architecture and the tile data format.

\subsection{Data Distribution}
To distribute our 3D stencil grid to the Tensix grid, we collapse the third dimension of the stencil onto the plane such that each Tensix core is allocated a column of tiles. For simplicity, we assume that the grid is evenly divisible by tiles in the horizontal plane.
Each core is allocated $N_z$ 64x16 tiles of the stencil where $N_z$ is the height of the 3D grid. We choose tiles of 64x16 instead of 32x32 to align better with the layout of tiles in memory, as discussed in \ref{sec:wormhole}, to facilitate the stencil data manipulation outlined below.


Under this layout, each core can update all local elements using only data from its four neighboring cores, in addition to vertical components which are local to the Tensix core. Figure \ref{fig:stencil} illustrates the 7-point stencil and how it maps to tiles and cores.

\begin{figure}[]
	\centering
	\subfigure[7-point stencil with tiles]{\includegraphics[width=.27\textwidth]{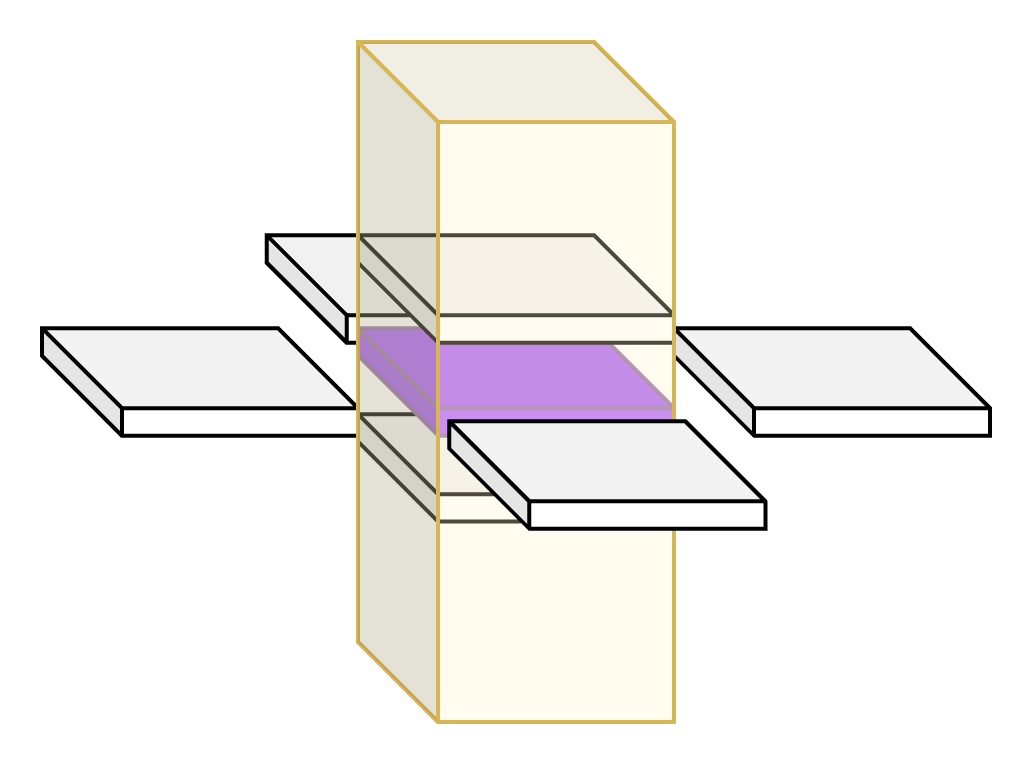}}
	\hfill
	\subfigure[Stencil from above for two grid point examples]{\includegraphics[width=0.27\textwidth]{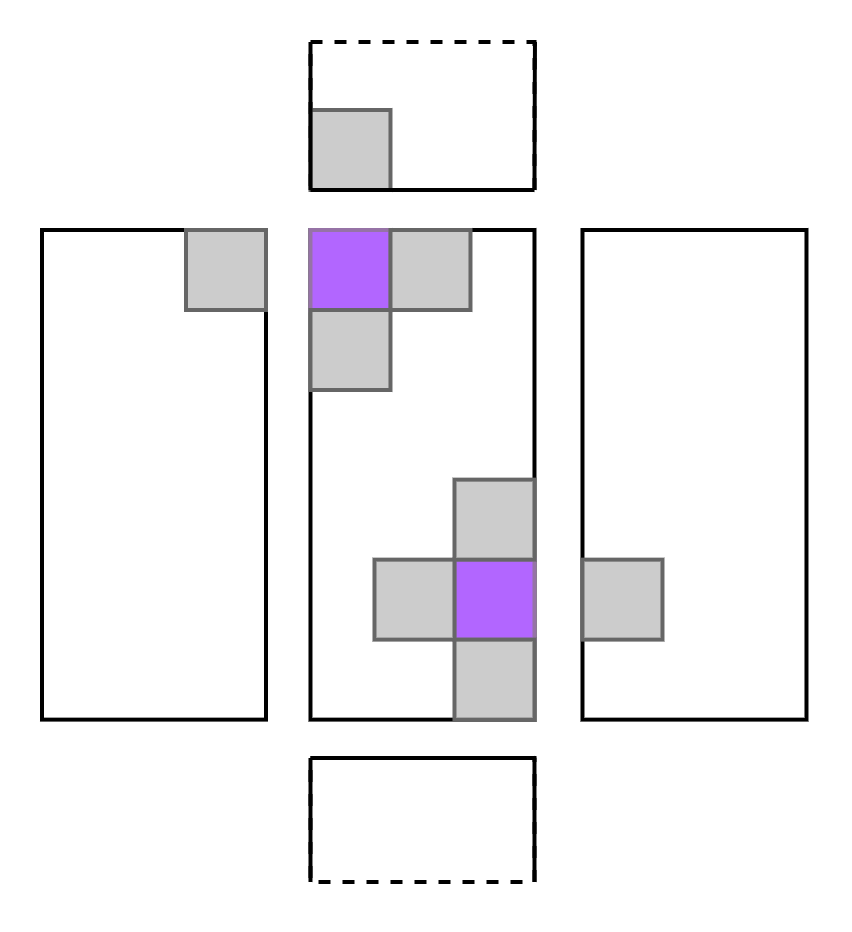}}
	\caption{Visualization of the 7-point stencil in Tenstorrent's tile-centric data format. The center point/tile is highlighted in purple.
		Figure (a) shows the stencil extended to a tile. The yellow column highlights the tiles that reside on the same Tensix core. Figure (b) shows the stencil from above, demonstrating how each element of a tile needs North, South,
		East, and West elements, which may reside on other tiles/cores.}
	\label{fig:stencil}
\end{figure}

The vertical stencil contributions can be computed easily by simply adding the center tile to the local above and below tiles. The horizontal stencil values, however, are more complicated because (1) they may reside on neighboring cores, and (2) they have different indices within the tile compared to the center element, so basic element-wise tile addition is insufficient. During every stencil operation, each core therefore performs both an exchange of all four boundary data with its four neighbors, and then the necessary shifting to align stencil elements and prepare for element-wise addition.

\subsection{Tile Shift for the Stencil Computation}
\label{sec:shift}

As mentioned, all computation on a Tensix core happens in SPMD fashion at the tile scale. In order to add two elements $a$ and $b$, the programmer needs to create two tiles, ensure that $a$ and $b$ are at the same position in each tile, and add the tiles together. The tile constraint particularly complicates the stencil computation.

For example, given a tile of elements $x_{i,j,k}$, the northern component of each element is $x_{i,j-1,k}$.
To add the northern element to the center, we therefore need to construct a tile which contains the $x_{i,j-1,k}$ elements at positions ${i,j,k}$. This is accomplished by ``shifting'' all the elements in the original tile down by one row, such that northern elements are now at coordinates that align with their corresponding center elements. Adding this shifted tile of northern elements to the center tile achieves one component of the stencil.




To implement the shift for each direction, we augmented the open source tt-metal API with a function that manually increments and decrements the read pointers of a circular buffer, similarly to previous work \cite{brown2024accelerating}.
Note that because tile pointers must align at 32B, pointer manipulation is restrained and pointers can only be incremented or decremented by multiples of 32B. Because of our choice of 64x16 element tiles, this maps to 1 row of a tile.

The northern elements can be produced by decrementing the read pointer by one row of elements (32B above the original tile). The southern tile can be generated in a similar manner (incrementing the pointer instead of decrementing). The eastern and western tiles require a shift left and right, which can't be achieved directly by pointer manipulation. Instead, this shift requires a tile transpose, followed by a shift up or down, incrementing the read pointer to get the eastern elements and decrementing to get the western elements. The shift completes with a second transpose to return the tile to its original orientation.
Once the shift operations are complete, the halo elements are filled in from neighboring cores via the NoC.

\subsection{Stencil Performance}

Figure \ref{fig:spmv-scaling} evaluates the weak scaling performance of the 7-point stencil. The stencil weak scales almost perfectly, except for additional costs incurred in the smallest configuration of a single core. Further performance analysis is performed by eliminating various components of the stencil to isolate the bottleneck.

First, we investigate the performance of the halo exchange: the ``no halo'' configuration in Figure \ref{fig:spmv-scaling} eliminates the halo exchange, and shows that the $1\times1$ and $2\times2$ grid configurations are unexpectedly expensive.
This overhead is isolated by the “no zero fill” results, which eliminate the code that sets boundary elements to zero.
Zero fill is expensive due to the high latency load and store access of the baby RISC-V's to the L1, and this overhead is more exposed in the small configurations.  In practice, the zero fill overhead is only significant for small Tensix core counts, which are unlikely to be used for real workloads.

The ``neither'' result removes both the halo and zero fill overheads, demonstrating that without these components, the stencil scales perfectly as expected.
The remaining runtime is largely dominated by the local stencil computation, which involves the tile shifts, transposes, and element-wise additions. This local compute is much more expensive than the communication, demonstrating the strength of the Wormhole NoC for stencil workloads. Future work could include a more detailed performance breakdown of the local compute alone, to isolate the overhead of tile transpose and shifts.
\begin{figure}[t]
	\centering
	\resizebox{0.35\textwidth}{!}{
		\begin{tikzpicture}
  \begin{axis}[
      xlabel=number of cores,
      ylabel=time/iter (ms),
      grid=both,
      ymin=.565,
      ymax=.6,
      legend style={at={(.6,0.2)}, anchor=west},
      mark size=2pt,
    ]
    \addplot [mark=triangle*, color=blue] table [x expr=\thisrow{x} * \thisrow{y}, y expr=((\thisrow{runtime})/ (\thisrow{iterations})),  col sep=comma] {MayaTaylor_Figs/fp32_sfpu/spmv_tracy.csv};
    \addlegendentry{total}

    \addplot [mark=triangle*, color=orange] table [x expr=\thisrow{x} * \thisrow{y}, y expr=((\thisrow{nohalo})/ (\thisrow{iterations})),  col sep=comma] {MayaTaylor_Figs/fp32_sfpu/spmv_tracy.csv};
    \addlegendentry{no halo}

    \addplot [mark=triangle*, color=green] table [x expr=\thisrow{x} * \thisrow{y}, y expr=((\thisrow{nofill})/ (\thisrow{iterations})),  col sep=comma] {MayaTaylor_Figs/fp32_sfpu/spmv_tracy.csv};
    \addlegendentry{no zero fill}

    \addplot [mark=triangle*, color=red] table [x expr=\thisrow{x} * \thisrow{y}, y expr=((\thisrow{neither})/ (\thisrow{iterations})),  col sep=comma] {MayaTaylor_Figs/fp32_sfpu/spmv_tracy.csv};
    \addlegendentry{neither}

  \end{axis}
\end{tikzpicture}}
	\caption{Weak scaling of the stencil. These results use 64 tiles per core and the FPU for BF16 operations. Results are averaged over 1000 iterations. }
	\label{fig:spmv-scaling}
\end{figure}

\section{Conjugate Gradient Implementation}
\label{sec:cg}
We combine the preceding kernels into a preconditioned conjugate gradient (PCG) solver for a 3D seven-point finite-difference stencil with zero Dirichlet boundary conditions. The sparse matrix-vector product is implemented matrix-free using the stencil of Section \ref{sec:stencil}, and the Jacobi preconditioner reduces to element-wise scaling. For simplicity, both are hardcoded in this proof-of-concept implementation; future work will consider more general sparse matrix representations.

We implement two variants of the PCG algorithm, one using BF16 and one using FP32. The BF16 implementation uses the FPU for all compute and keeps the residual norm device-resident in SRAM between iterations.

The FP32 implementation uses the SFPU (required for this higher precision) and a split-kernel approach, where the residual norm is written back to the host at every iteration. This difference reflects the added programming difficulty of the SFPU, which favors the simpler split-kernel implementation. While the split-kernel approach does incur additional host-synchronization overhead, we expect the majority of the FP32 slowdown to instead stem from the higher cost of SFPU operations relative to the FPU (Section \ref{sec:arithmetic}); disentangling the two remains future work.

We validate correctness and convergence against a CPU reference implementation; a detailed characterization of BF16 numerical stability (e.g., residual behavior, sensitivity to overflow/cancellation) is left to future work.

\subsection{Evaluation}

\begin{figure}[]
	\centering
	\subfigure[Strong scaling of FP32 and BF16 implementation with fixed problem size of 64x16 tiles and 164x4 tiles respectively, to maximize problem size at 2x2 cores.]{
		\resizebox{0.35\textwidth}{!}{
			\begin{tikzpicture}
  \begin{axis}[
      xmode=log,
      ymode=log,
      xlabel=number of cores,
      ylabel=time/iter (ms),
      grid=both,
      ymin=.05,
      ymax=4,
      ytick={.1, .2, .4, .8,  1.6, 3.2},
      yticklabels={.1, .2, .4, .8, 1.6, 3.2},
      xtick={1,2,4,8,16,32,64},
      xticklabels={1,2,4,8,16,32,64},
      minor y tick num=4,
      mark size=2pt,
      legend style={at={(0.2,0.15)}, anchor=south, legend columns=1},
    ]
    \addplot [color=red] table [x expr=\thisrow{corex} * \thisrow{corey}, y expr=(\thisrow{ideal} / \thisrow{iterations}),  col sep=comma] {MayaTaylor_Figs/scaling/strong-scaling-host-timing.csv};

    \addplot [mark=square*, color=green!70!black] table [x expr=\thisrow{corex} * \thisrow{corey}, y expr=(\thisrow{total} / \thisrow{iterations}),  col sep=comma] {MayaTaylor_Figs/scaling/strong-scaling-host-timing.csv};
    \addlegendentry{BF16}

    \addplot [mark=triangle*, color=blue] table [x expr=\thisrow{corex} * \thisrow{corey}, y expr=(\thisrow{remaining} / (\thisrow{iterations} - 1)),  col sep=comma] {MayaTaylor_Figs/fp32_sfpu/strong-scaling.csv};
    \addlegendentry{FP32}

    \addplot [color=red] table [x expr=\thisrow{corex} * \thisrow{corey}, y expr=(\thisrow{ideal} / (\thisrow{iterations} - 1)),  col sep=comma] {MayaTaylor_Figs/fp32_sfpu/strong-scaling.csv};
    \addlegendentry{ideal}
  \end{axis}
\end{tikzpicture}}}
	\subfigure[Weak scaling at max problem size for FP32 (64 tiles, or 167,936 elements of $x$ per core) and BF16 (164 tiles, or 671,744 elements of $x$ per core), normalized per tile.]{
		\resizebox{0.35\textwidth}{!}{
			\begin{tikzpicture}
  \begin{axis}[
      xlabel=number of cores,
      ylabel=time/iter per tile (ms),
      grid=both,
      ymin=0,
      ymax=.05,
      legend style={at={(.7,.55)}, anchor=west},
      mark size=2pt,
    ]
    \addplot [mark=triangle*, color=blue] table [x expr=\thisrow{x} * \thisrow{y}, y expr=((\thisrow{run1} + \thisrow{run2} + \thisrow{run3})/ (\thisrow{iterations} - 1) / 3 / 64),  col sep=comma] {MayaTaylor_Figs/fp32_sfpu/wormhole-weak.csv};
    \addlegendentry{FP32}

    \addplot [mark=square*, color=green!70!black] table [x expr=\thisrow{corex} * \thisrow{corey}, y expr=((\thisrow{total})/ \thisrow{iterations} / 164),  col sep=comma] {MayaTaylor_Figs/scaling/bf16-weak-164tiles.csv};
    \addlegendentry{BF16}

    \addplot [color=red] table [x expr=\thisrow{corex} * \thisrow{corey}, y expr=((\thisrow{ideal})/ \thisrow{iterations} / 164),  col sep=comma] {MayaTaylor_Figs/scaling/bf16-weak-164tiles.csv};

    \addplot [color=red] table [x expr=\thisrow{x} * \thisrow{y}, y expr=((\thisrow{ideal})/ (\thisrow{iterations} - 1) / 64),  col sep=comma] {MayaTaylor_Figs/fp32_sfpu/wormhole-weak.csv};
    \addlegendentry{ideal}
  \end{axis}
\end{tikzpicture}}}
	\caption{Scaling results for the CG solve on Tenstorrent Wormhole. Results are produced via host-side timing of the entire solve, averaged over 100 iterations.}
	\label{fig:scaling}
\end{figure}

\begin{table*}
	\centering
	\small
	\begin{tabular}{lcccc}
		\toprule
		\textbf{Specification} & \textbf{n150d} & \textbf{n300d} & \textbf{H100} \\
		\midrule
		Vendor                 & Tenstorrent    & Tenstorrent    & Nvidia        \\
		TDP (W)                & 160            & 300            & 350           \\
		Manufacturing Node     & GF 12nm        & GF 12nm        & TSMC N4       \\
		Peak Mem BW (GB/s)     & 288            & 576            & 3,900         \\
		Memory                 & 12 GB GDDR6    & 24 GB GDDR6    & 80 GB HBM3    \\
	\end{tabular}

	\vspace{0.05in}
	\caption{
		High-level architectural characteristics of the Tenstorrent Wormhole n300d PCIe accelerators, Nvidia H100 GPU.
		The n150d is provided for reference, as it is a single-die version of the n300d.
	}
	\label{tab:hpc}
\end{table*}

The PCG algorithm can run on any configuration of Tensix cores, allowing us to scale compute resources by varying the size of the Tensix sub-grid used.
Although our test system contains an n300d, all evaluations use only a single Tensix die and up to an $8 \times 7$ sub-grid. 
The maximum problem size is prescribed by the memory capacity of each core's local SRAM, which is approximately 1.5MB. After accounting for stack, program storage, and circular buffers, the maximum input size per core is 64 tiles for the FP32 split-kernel implementation and 164 tiles for the BF16 implementation, or global problem sizes of $3.6$ million and $9.4$ million elements respectively. We do not evaluate performance on larger problem sizes which require spill-over into DRAM, and future work should explore the feasibility of overlapping DRAM streaming access with PE-local operations.

Figure \ref{fig:scaling} shows strong and weak scaling results for both implementations of PCG on the Tenstorrent Wormhole. The FPU scales well in both cases and the SFPU exhibits good weak scaling but more irregular strong scaling. Notably, the SFPU implementation is approximately 2 times slower than the FPU implementation when normalized against the problem size, due to the increased cost of the SFPU operations.

Table \ref{table:gpu-comp} compares the performance of the Wormhole against a 32-bit H100 implementation of PCG. The H100 (Tab.~\ref{tab:hpc}) results were gathered under gcc 13.3.0, CUDA 12.8.1, and Kokkos 4.6.2.
with CG assembled from four individual kernels: \texttt{norm}, \texttt{dot}, \texttt{axpy}, and \texttt{SpMV}.
Rather than a custom structured SpMV kernel, our implementation leverages Nvidia's cuSPARSE sparse linear algebra library.
cuSPARSE does not provide a structured matrix format, so the structure was realized via the Sliced ELL format, generally recognized as state-of-the-art in performance for matrices with limited row-length variability.
The \texttt{norm}, \texttt{dot}, \texttt{axpy} kernels are implemented via the Kokkos programming system in a naive, straightforward way. The subcomponent times are measured by wrapping the relevant operations in \texttt{cudaEventRecord} calls, followed by calls to \texttt{cudaEventSynchronize} and \texttt{cudaEvent\-ElapsedTime}.
Due to the implementation of \texttt{dot} via Kokkos' \texttt{paral\-lel\_reduce}, the \texttt{dot} time includes transferring the residual norm back to the host.

To provide context for these results, Table~\ref{tab:hpc} summarizes some high-level architectural characteristics of the Tenstorrent Wormhole PCIe accelerator and the Nvidia H100 GPU. Because we are only utilizing one Tensix die of the n300d,
the specs of the n150d are more relevant for comparison, especially in terms of TDP, manufacturing process, and memory bandwidth.
The H100 outperforms Wormhole in raw time per iteration, with Wormhole being approximately 4.3x slower in BF16 and 8.8x slower in FP32. Nevertheless, because our experiments use only a single 12nm Wormhole die with far lower available memory bandwidth and TDP than the 4nm H100, the BF16 result suggests a narrower practical competitiveness gap than raw runtime alone would imply.
Future work will explore full utilization of the n300d and direct performance-per-watt comparisons with H100 using measured power rather than TDP as a proxy.


\begin{table}[]
	\centering
	\small

	\begin{tabular}{ll}
		\toprule
		\textbf{Implementation} & \textbf{Time/Iteration (ms)} \\
		\midrule
		H100                    & 0.28                         \\
		Wormhole BF16           & 1.20                         \\
		Wormhole FP32           & 2.45                         \\
		\bottomrule
	\end{tabular}

	\vspace{0.05in}
	\caption{
		H100 and Wormhole performance comparison for PCG for $512\times112\times64$ grid. The Wormhole implementations are evaluated on an 8x7 grid of Tensix cores with 64 tiles per core.
	}
	\label{table:gpu-comp}
\end{table}

\section{Conclusion}
\label{sec:conclusion}

In this work, we have presented implementations of three numerical kernels as well as two implementations of the preconditioned conjugate gradient (PCG) method using a seven-point finite difference stencil.
Our PCG implementation represents the most complex numerical algorithm successfully deployed on Wormhole to date, incorporating architecture-specific optimizations tailored for sparse numerical computations.
The key achievements of this work include:

\begin{itemize}
	\item Successfully extending the computational complexity boundary of an implementation on the Wormhole accelerator beyond traditional neural network workloads.
	\item Demonstrating performance within $8.8 \times$ of an H100 baseline.
\end{itemize}

While spatial architectures like Wormhole show promise for numerical computing, significant challenges remain in terms of programmability and generalization. The explicit management of data movement and parallelization, while enabling high performance for regular workloads, presents substantial complexity for irregular algorithms.

Future work on the Wormhole architecture should explore more general sparse matrix representations, multi-device scaling, and the development of higher-level programming abstractions to improve accessibility and usability for a broader range of scientific computing applications.
Extending this work to additional numerical methods such as multigrid solvers, FFTs and irregular graph algorithms would help establish the broader applicability of spatial architectures for scientific computing and identify common optimization patterns that could inform higher-level programming abstractions.

In light of the broad potential of spatial architectures for scientific computing, this work provides a foundational demonstration of traditional HPC algorithms on emerging AI-accelerator architectures. The results presented here suggest that with careful consideration of data movement and parallelization, spatial architectures can be effectively leveraged for a range of numerical methods beyond their original design intent.

\section{Acknowledgements}
We would like to thank the Sandia advanced architectures testbed program for providing computing resources for this publication.
Additionally, would like to thank Felix LeClair, Nikola
Cvetković, and Neil Sexton of Tenstorrent for their frequent technical discussions.

This report was prepared as an account of work sponsored by an agency of the United States Government. Neither the United States Government nor any agency thereof, nor any of their employees, makes any warranty, express or implied, or assumes any legal liability or responsibility for the accuracy, completeness, or usefulness of any information, apparatus, product, or process disclosed, or represents that its use would not infringe privately owned rights. Reference herein to any specific commercial product, process, or service by trade name, trademark, manufacturer, or otherwise does not necessarily constitute or imply its endorsement, recommendation, or favoring by the United States Government or any agency thereof. The views and opinions of authors expressed herein do not necessarily state or reflect those of the United States Government or any agency thereof.

\bibliographystyle{plain}
\bibliography{MayaTaylor}

\end{document}